\documentclass{kluwer}
\usepackage[dvips]{graphicx}
\usepackage{epsfig}
\begin{document}
\begin{article}
\begin{opening}
\title{OPTIMA: A Photon Counting High-Speed Photometer}

\author{C. \surname{Straubmeier$^{1,2}$}\email{cstraubm@ph1.uni-koeln.de}}
\author{G. \surname{Kanbach$^{1}$}\email{gok@mpe.mpg.de}}
\author{F. \surname{Schrey$^{1}$}\email{fzs@mpe.mpg.de}}
\institute{$^{1}$Max-Planck-Institut f\"ur extraterrestrische Physik, Garching\\
$^{2}$I. Physikalisches Institut, University of Cologne}

\begin{abstract}
OPTIMA is a small, versatile high-speed photometer which is primarily intended 
for time resolved observations of young high energy pulsars at optical 
wavelengths.
The detector system consists of eight fiber fed photon counters based on avalanche 
photodiodes, a GPS timing receiver, an integrating CCD camera to ensure the 
correct pointing of the telescope and a computerized control unit.
Since January 1999 OPTIMA proves its scientific potential by measuring a very 
detailed lightcurve of the Crab Pulsar as well as by observing cataclysmic variable
stars on very short timescales.

In this article we describe the design of the detector system focussing on the
photon counting units and the software control which correlates the detected
photons with the GPS timing signal.
\end{abstract}

\abbreviations{
\abbrev{OPTIMA}{Optical Pulsar Timing Analyser}
\abbrev{APD}{Avalanche Photodiode}
\abbrev{CCD}{Charge Coupled Device}
\abbrev{GPS}{Global Positioning System}
\abbrev{PMT}{Photomultiplier Tube}
\abbrev{UTC}{Universal Time Coordinated}
}

\end{opening}

\section{Introduction}

Considering pulsar detections throughout the electromagnetical spectrum from radio to 
$\gamma$-rays, the low number of indisputable identifications in the optical band is 
striking.
Up to now the search for optical emissions modulated at the pulsars' rotational 
frequencies has only succeeded in five significant or at least suspected detections 
(Cocke et al. 1969, Wallace et al. 1977, Middleditch \& Pennypacker 1985, 
Shearer et al. 1997, Shearer et al. 1998).
Nevertheless the optical emission from the neutron star's surface and magnetosphere
can provide important insights into the emission processes and thermal conditions 
of pulsars. 
In the magnetosphere it marks the onset of nonthermal high energy processes that 
extend to X- and $\gamma$-ray frequencies and, from the hot surface as the Rayleigh 
Jeans part of a thermal spectrum, it provides estimates on the size of the star 
and its equation of state.
To gain more information on pulsar emission in this decisive wavelength range, 
the $\gamma$-ray group of the Max-Planck-Institut f\"ur extraterrestrische Physik 
decided in 1996 to build a new high-speed photometer, the Optical Pulsar Timing 
Analyser (OPTIMA), which should be useable as a stand-alone guest instrument at 
various observatories.

Since January 1999, more than one year before its final completion in fall 2000,
the OPTIMA detector system has been used at several international telescopes to 
fine-tune its operation and to acquire first scientific data.
During this commissioning phase we observed amongst others the well known Crab 
pulsar to demonstrate the very high time resolution of a few microseconds and its 
long-term stability over more than three consecutive days \cite{Straubmeier Thesis}.

With the completion of the high-speed photometer in fall 2000 the OPTIMA detector 
is now ready for scientific useage on a wide variety of astronomical sources 
displaying variations of the intensity of their optical signal on short timescales.

\section{Principle of Operation}

The operation of the OPTIMA high-speed photometer is based on photon-counting, 
recording the arrival-time of every detected photon with a precision of a few 
microseconds. 
As the visual flux from most known or suspected optical pulsars is very low, a direct 
frequency analysis or stroboscopic observation of the measured intensity is 
impractical and precludes the later use of slightly different pulsar ephemerides or 
the derivation of longer term variations.
On the other hand the individual recorded photons allow the use of rotational periods 
derived from other wavelength bands at a later stage of offline data analysis. 
The acquired arrival-times, properly corrected to the solar system barycenter, can 
then be converted to rotational phases by folding and a phase-coherent lightcurve 
histogram can be derived.
Nevertheless the recording of the unprocessed arrival-times during the observation 
enables us further to search for pulsations at unknown frequencies in sources with 
a sufficiently high flux-level and to measure occultations, outbursts or 
quasi-periodic oscillations of other targets, like compact binary systems.

\section{Hardware Layout}

In order to use OPTIMA as a guest instrument on various large observatories, a basic 
requirement on the design of the detector system was, that it should be easily 
adaptable to different telescope configurations.
As a consequence OPTIMA is a completely autonomous system incorporating all optics, 
electronics and computers which are neccessary for its operation, so that only minor 
optical and mechanical adjustments are required to match the focal scale of the used 
telescope.   
In the course of its development, OPTIMA has been used on the Cassegrain foci of the 
Mt. Stromlo 74inch, ESO/MPG/La Silla 2.2m, Skinakas/Crete 1.3m and primarily on the 
3.5m telescope of the German-Spanish Observatory on Calar Alto, Spain.

\begin{figure}[thp]
\centering
\caption{Simplified sketch of the hardware layout of OPTIMA}
\label{Sketch}
\epsfig{file=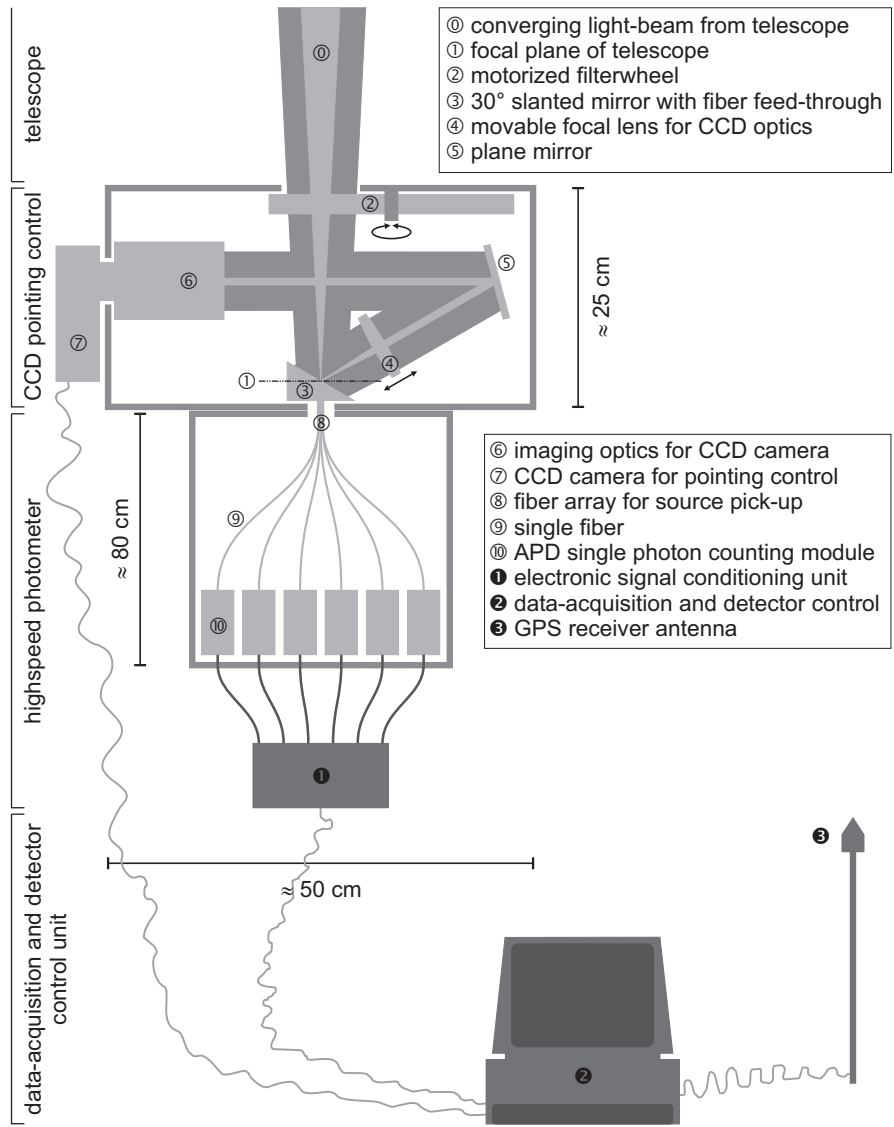, width=\textwidth}
\end{figure}

The detector hardware consists basically of four subsystems, which are sketched in 
figure~\ref{Sketch} and will be discussed in turn in the next sections of this 
article: 

\begin{itemize}
\item{The high-speed APD counters}
\item{The GPS receiver and high frequency time base}
\item{Two personal computers for data-acquisition and detector control}
\item{The integrating CCD camera for target acquisition and pointing verification}
\end{itemize}

\subsection{High-Speed Photometer}

The basic purpose of a high-speed photometer is to record the radiation flux of a given 
source with a time resolution in the range of milliseconds or less.
In the case of OPTIMA, which is primarily designed for studying the optical 
lightcurves of extremely faint pulsars (m$_V$ $\approx$ 25$^{m}$), the observed flux 
of the astronomical target lies considerably below the brightness of the night sky 
(m$_V$ $\approx$ 22$^{m}$ arcsec$^{-2}$) at even the best astronomical sites.
To minimize the degradation of the source's signal due to the underlying atmospheric 
background, the flux of the target is isolated in the focal plane of the telescope by
the use of an optical fiber pick-up which acts as a diaphragm.
The signal to background ratio can then be maximized by matching the diameter of the 
fiber to the point spread function of the telescope and the expected seeing conditions 
(see section~\ref{FiberPickUp}). 

On a three meter class telescope during a typical pulsar observation in a moon-less 
night, the average detected photon-rate is about 1000 Hz, with more than 99\% of it  
due to atmospheric background.

As the emission from optical pulsars is normally strongly polarized (Chen et al. 1996,
Graham-Smith et al. 1996), a possible instrumental polarization may significantly 
affect the shape and dynamics of the observed lightcurve.
To rule out such systematic effects in measurements with the OPTIMA high-speed 
photometer, a series of tests with unpolarized as well as linearly polarized 
calibration lamps have been performed in the laboratory.
By equipping the motorized filterwheel of OPTIMA (item no. 2 in figure~\ref{Sketch}) 
with a continuously rotating polarization filter, the expected sinusoidal modulation 
of a polarized lightsource was clearly detected.
When using an unpolarized input source no such modulation was found at the 
rotational frequency of the filterwheel.

This behaviour proves that no instrumental polarization is present and we therefore 
conclude that whithout additional filters the OPTIMA high-speed photometer is 
insensitive to the polarization properties of the observed source. 

\subsubsection{Fiber Pick-Up}
\label{FiberPickUp}

\begin{figure}[thb]
\centering
\caption{Hexagonal bundle of seven optical fibers fed through a hole in the slanted 
mirror at the focal plane of the telescope.
For handling reasons the fibers are packed into a protective steel tube.
The horizontal bright line is formed by an auxiliary mark on the mirror to ease 
focussing of the CCD optics.}
\label{Fibers}
\epsfig{file=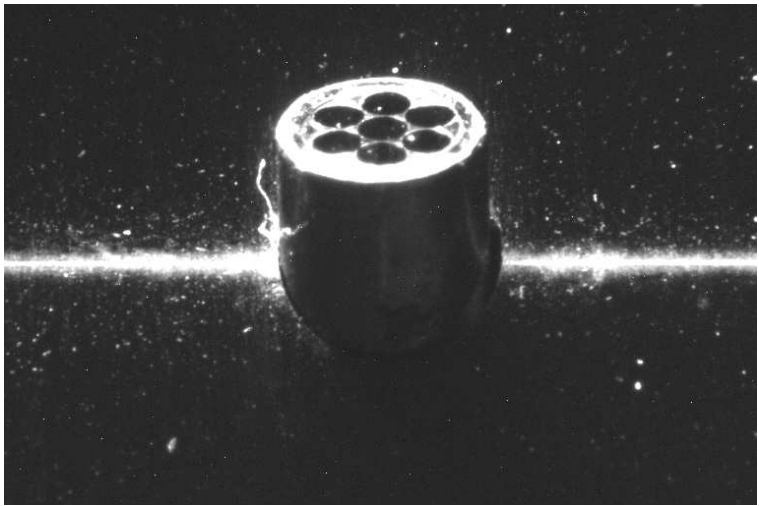, width=100mm}
\end{figure}

In order to minimize the contribution of the background sky and to maximize the 
signal to background ratio, the optical flux of the observed source and its underlying 
and adjacent sky region is picked up at the focal plane of the telescope by a hexagonal 
bundle of seven optical fibers (see figure~\ref{Fibers}).
An additional single fiber is placed at a typical distance of about one arcminute 
from this bundle to monitor the sky brightness.

By the use of exchangeable fiber tapers (optical fibers having different radii at both 
ends) it is possible to match the fiber diameter in the focal plane to the focal scale
of the used telescope and the expected seeing conditions at the observing site.
Usually a diameter corresponding to about two seeing discs is chosen.
Having the target precisely positioned on the central fiber, this setup concentrates more 
than 98\% of the flux of the observed object in this detector channel with the least 
possible amount of background light from the earth's atmosphere.
In addition this setup leaves enough room to allow for small pointing variations being 
present at every telescope. 

The recorded signals of the six fibers around the target and the more distant single 
fiber which monitor the sky background are used to develop a time-dependent model 
of this background intensity at the location of the source.
Thus it is possible to derive the time resolved flux of the source without contribution 
of the background radiation, which is a vital step for example to determine the extent 
of time invariant or only slowly variable emission of the observed object.   

On the other end of each fiber the photons are fed to an APD detector, which provides 
the necessary conversion into electrical signals at a very high quantum efficiency and 
a very short response time.

\subsubsection{Avalanche Photodiodes}

In order to obtain a statistically significant measurement of faint sources within a 
limited observing time it is very important to convert the highest possible fraction 
of incoming photons into electronic countable signals.
The crucial parameter for this conversion is the wavelength dependent quantum 
efficiency of the used photodetector.
Up to now the standard detectors for recording single optical photons with time 
resolutions of a few microseconds are photomultiplier tubes (PMTs).
PMTs deliver a high signal to noise ratio for the electrical pulse of a detected 
photon but the standard photo-cathodes have only a comparably low peak quantum 
efficieny of about 20\% and a narrow wavelength range of sensitivity.
Similar properties apply for the photo-cathodes of the Multi Anode Microchannel
Array (MAMA) detectors used by Shearer et al., 1997, 1998.
Cryogenic detectors based on superconducting tunnel junctions (Perryman et al. 1999,
Rando et al. 2000) are also still limited to system efficiencies of around 
30\%, although they offer intrinsic spectral resolution. 
The cryogenic transition edge sensor developed by Romani et al., 1999, achieves 
only 2\% system efficiency so far.

To overcome these limitations OPTIMA is based on state-of-the-art Avalanche Photodiodes
(APDs).
These new silicon based semiconductor devices offer a peak quantum efficiency of more
than 60\% and a wide band of sensitivity ranging from 400 to 1050 nanometers.

\begin{figure}[thb]
\centering
\caption{Quantum Efficieny of an APD Single Photon Counting Module based on 
data from the manufacturer, PerkinElmer Inc.}
\label{SPCMQE}
\epsfig{file=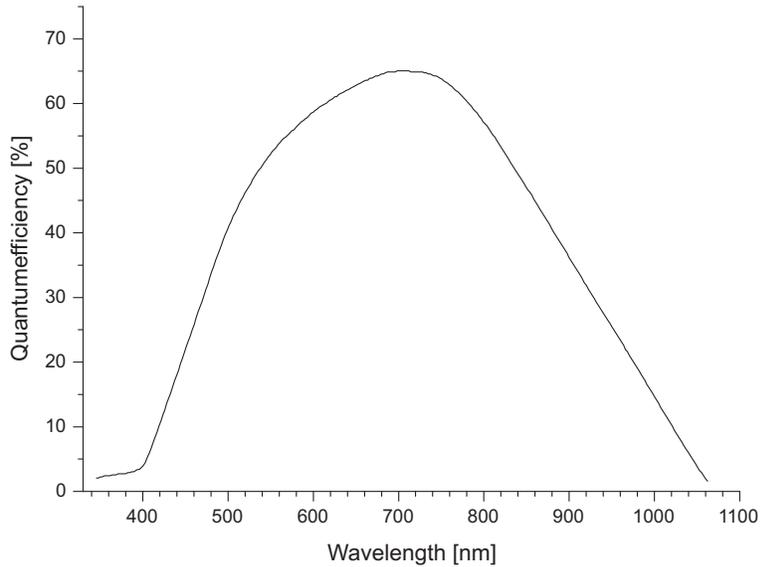, width=100mm}
\end{figure}

These two features result in a broadband gain of sensitivity of a factor of six 
in comparison to standard photomultiplier tubes.
It is technically demanding to operate APDs as analog devices in a low noise and 
time-stable single photon counting configuration.
The internal gain close to avalanche break-through critically depends on the 
temperature and the applied bias voltage. 
As a consequence the detectors and their preamplifiers have to be cooled to about 
-30$^\circ$C and controlled to a fraction of a degree by a multistage Peltier system. 
The needed bias voltage in the range of several hundred volts needs to be adjusted 
and precisely maintained within a few millivolts.

To speed up the completion of the OPTIMA detector system and to keep the detector 
electronics simple, we decided not to continue the initial development of a multi 
channel analog APD array for OPTIMA but to use eight units of the commercial available 
APD based Single Photon Counting Modules of PerkinElmer Inc. instead.
These highly integrated devices operate in a Geiger counter mode where a photon 
initiated avalanche pulse is quenched by the instantaneous reduction of the bias
voltage. 
The used APDs are of 0.2 mm diameter and are electrically cooled. 
The detector units selected for OPTIMA offer low dark count rates of typically less 
than 30 Hz, are insensitive to electromagnetic interference and are very reliable in 
operation. 
They can operate up to photon-rates of several 100 kHz before noticeable deadtime 
losses occur.

\subsection{GPS Timing}

To measure the lightcurves of pulsars with typical rotation periods from 33 ms (Crab) 
to 237 ms (Geminga) with a photon counting system like OPTIMA, the arrival-times of 
the individual detected photons have to be recorded with the precision of a few 
microseconds.
This accuracy must be maintained continuously for the whole observing time spent on 
a selected target, which in the case of very faint sources might even extend over 
several consecutive nights.
Otherwise the periodic signal of a weak source would be smeared out during the 
process of folding the recorded arrival-times with the rotational period of the 
pulsar.

Keeping in mind that OPTIMA should be usable at different telescopes around the 
world, the signals of the global GPS satellites are the best available absolute 
time base.
Using a special receiver which can process the clock pulses of up to six satellites
simultaneously, it is possible to reach an absolute time accuracy of the one 
pulse per second GPS clock signal of better than two microseconds.
Using this periodic signal to discipline a high frequency oscillator, a high 
resolution time base can be established with a maximum time deviation given by the 
error of the GPS clock signal.
For OPTIMA we use a GPS based time and frequency processor of Datum Inc. which 
provides a continuous UTC time signal with a self adjusting absolute accuracy of 
better than two microseconds to the system bus of a personal computer as well as 
on external connectors.   

\subsection{Data-Acquisition}
\label{Dataacquisition}

The most crucial task of the computer based data-acquisition unit is to correlate 
the electronic signals of the APD detector modules with the high resolution time base 
and thus assign UTC arrival-times to each detected photon.
This association is done on hardware level to ensure a reliable operation even on a 
non realtime operating system and under high system load.
The timing of the conversion cylces of the data-acquisition card is therefore done  
by the GPS based high frequency oscillator, so that the transfer of the APD detector 
signals to the computer system is running at a fixed rate.

As the precise starting time of each software triggered acquisition sequence is well 
known, the remaining job of the controling software is to count the number of 
conversion cycles since the start of the sequence and to store this sequential 
number together with an identifier of the respective detector channel for each 
detected photon.
Based on the cycle number, the acquisition frequency and the absolute time of the 
start of the sequence the UTC arrival-time of every recorded photon can be restored  
during data analysis.

If one considers observing times of several nights, eight detector channels, an 
acquisition frequency of several 100 kHz and an atmospheric background photon-rate 
of about 1 kHz the total amount of data is quite formidable. 
However the chosen method of storing the arrival-time data poses quite acceptable 
demands on the required computer memory because all conversion cycles without any 
detected photon can be skipped and most of the stored data consists of small
integer numbers. 
Nevertheless one night of observation still produces several gigabytes of data.

\subsection{CCD Positioning Camera}

By the use of optical fibers only the flux of the targeted object and a small 
adjacent region of sky is transfered from the focal plane of the telescope to the 
APD high-speed counters.
The light of the more extended region of surrounding sky (about 3-4 arcminutes in 
size) is reflected by a slanted mirror into a focal reducer and imaged by a ST-7 
integrating CCD camera of the Santa Barbara Instrument Group.
In conjunction with this CCD image the catalogued astronomical positions of several 
bright stellar objects in the vicinity of the target can be used to verify the 
acquisition of the fiber array and the correct pointing of the telescope.
This ability to point the telescope using bright guide stars with known separations 
from the target makes it possible to position the high-speed photometer even on 
very weak sources, which themselves are too faint to be visible within a reasonable 
integration time.

With appropriate electrical connections to the telescope control system the OPTIMA 
CCD camera can also operate as a guiding system to ensures the long-term stable 
pointing of the telescope by automatically correcting small deviations.
This possibility enables us to use OPTIMA at telescopes which do not provide
a stationary stand-alone star-tracker for guest instruments. 

\section{Control Software}

Most of the functions of the OPTIMA detector system are remotely controlled by 
software running on two personal computers located at the observers' room.
From these two machines (one for the control and the data-acquisition of the 
high-speed photometer and one for the control of the integrating CCD camera) all 
observing parameters of OPTIMA can be monitored and changed interactively.
 
In contrast to the readout and image-processing software for the CCD camera, where 
a well standardized CCD command language is available and several commercial 
software products can be used, most of the required software for the operation of 
the high-speed photometer was newly developed by ourselves.
The software includes operation modes for data-acquisition, setup and control of the 
APD detector modules and the GPS receiver as well as some low level data analysis 
procedures including a flexible data interface to export the pre-processed data to 
other scientific software packages.
The analysis part especially provides some fundamental tasks for pulsar research 
like the correction of the recorded local UTC arrival-times to the reference frame 
of the solar system barycenter, the calculation of phase-coherent intensity 
histograms based on externally supplied rotational parameters of the observed source, 
or a variety of frequency analysis tools for the search for pulsations at unknown 
periods.

The process of barycentering the recorded UTC arrival-times corrects the acquired 
data for timing shifts and Doppler effects due to the daily and annual motion of the 
earth and makes the individual measurements comparable between different epochs and 
observing sites.
The possibility to use the known ephemerides of the source from observations in other 
wavelength bands allows the calculation of lightcurve histograms even of very faint 
objects, where the optical periodic signal is too weak to determine the frequency 
of pulsation with sufficient precision from the OPTIMA data alone.  

\section{First Light of the OPTIMA Detector on the Crab Pulsar}

To start with the commissioning and fine tuning of the detector system as soon as
possible, OPTIMA was designed to be already useable for first astronomical 
observations with a smaller number of detector channels than the final eight APDs.
In late December 1999, when the assembly and lab testing of OPTIMA with two detector 
channels was successfully completed, we observed the bright Crab Pulsar 
(PSR~B0531+21, m$_V =~$16.6$^{m}$) at the 3.5m telecope of the German-Spanish 
observatory on Calar Alto, Spain.
Because of its relatively high visual brightness and high rotational frequency, 
this pulsar has been a favourite target for the demonstration of the respective 
capabilities of many optical high-speed photometers (Beskin et al. 1983, 
Eikenberry et al. 1997, Golden et al. 2000, Perryman et al. 1999, Romani et al. 1999).   
Considering the limiting properties of the system due to the small number of 
available detector channels and due to a not yet optimized set of optical fibers, 
the results of these short observations nevertheless give a striking impression of 
the performance and future potential of our new high-speed photometer.  

\begin{figure}[t]
\centering
\caption{Lightcurve of the Crab Pulsar (PSR B0531+21) as measured with OPTIMA in 
December 1999 at the Calar Alto 3.5m telescope in a 10 minute exposure.  
The plotted numbers represent the absolute countrates of the detector, which are 
neither corrected for the contribution of the atmospheric background radiation 
nor for the intensity of the emission nebula surrounding and overlapping the pulsar. 
The horizontal line at approximately 4000 Hz shows the result of a linear fit to the phase 
interval of lowest intensity at 0.7729 $\leq$ $\varphi$ $\leq$ 0.8446.
This level roughly identifies the flux of the nebula at the position of the pulsar 
(Percival et al. 1993)(Straubmeier 2001).
At the epoch of this observation one rotational cycle of the pulsar equals 33.505115 
msec and the size of the plotted phase intervals is approximately 112 $\mu$sec 
(300 bins per rotation). 
For clarity two consecutive cycles of pulsation are plotted.}
\label{CrabLC}
\epsfig{file=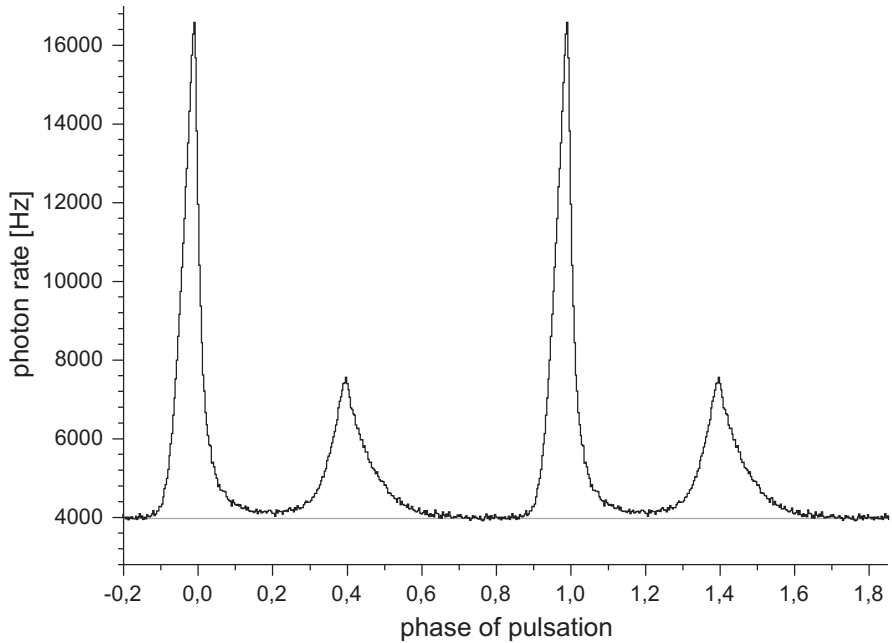, width=\textwidth}
\end{figure}

Using the well known periodicity of the Crab Pulsar as one of the most precise clocks 
available and analysing the measured phase positions of the sharp main peak of the 
lightcurve on four consecutive nights, the absolute timing of OPTIMA proved to be 
accurate to better than 100 $\mu$sec over an observing time span of more than 80 
hours. 
At the highest possible time resolution the analysis was limited by the low number of 
photons recorded from the source, which was observed for only about 20 minutes 
during each night, resulting in a total exposure time of about 80 minutes.
The given upper limit of the timing accuracy of OPTIMA is therefore expected to be 
improved by about one order of magnitude by a more extensive observation of this 
source, which is being planned for the near future.         

\subsection{Phase-Coherent Folding of Arrival-Times}

In figure \ref{CrabLC} we show the lightcurve of the Crab Pulsar as measured with 
OPTIMA in December 1999. 
For the calculation of this intensity histogram the recorded arrival-times of the 
detected photons have been corrected to the solar system barycenter and were folded 
with the rotational period of the source as published monthly by the Jodrell Bank 
Radio Observatory \cite{Crab Ephemeris}.
With only two detector channels available it is not possible to isolate the optical 
flux of the pulsar from the contribution of the underlying spatially highly 
inhomogenous emission of the Crab Nebula and the night sky background.
The plotted counting rates therefore represent the summed intensity of these 
components per phase interval. 
For clarity figure \ref{CrabLC} shows two rotational cycles of the pulsar with one 
full rotation corresponding to 33.505115 msec. 
The zero point of the plotted phase index is defined by the arrival of the main radio 
pulse at the solar system barycenter.

A comparison of the basic characteristics of our measured lightcurve to the published 
data from the former high-speed photometer aboard the Hubble Space Telescope 
\cite{Percival1993} shows that the OPTIMA detector system contributes no detectable 
timing noise or non-linear intensity response to the optical signal variation of the 
pulsar \cite{Straubmeier Thesis}.
The obtained lightcurve demonstrates that OPTIMA can resolve even large intensity 
variations down to a timescale of fractions of milliseconds and presumably even 
further when more observational data becomes available. 

\subsection{Frequency Analysis}

For the detection of periodic signals with very sharp features covering only a small 
fraction of the full period the Z$^{2}$ test statistics of Buccheri \cite{Buccheri1983} 
proved to be very sensitive.
In addition to the fundamental mode of oscillation this test considers an adjustable 
number of higher harmonics as well and therefore reaches a high sensitivity for 
non-sinusoidal signals.
Regarding the overall shape of the lightcurve of the Crab Pulsar with a phase extent of 
pulsed emission of approximately 20\% we chose a Z$^{2}_{10}$ test, which includes 
the fundamental mode and its first nine harmonics.

\begin{figure}[h]
\centering
\caption{Frequency spectrum of the Crab Pulsar in December 1999.
The plotted intensities are the results of a Z$^{2}_{10}$ test considering the 
fundamental mode of pulsation and its first nine harmonics.} 
\label{CrabPScan}
\epsfig{file=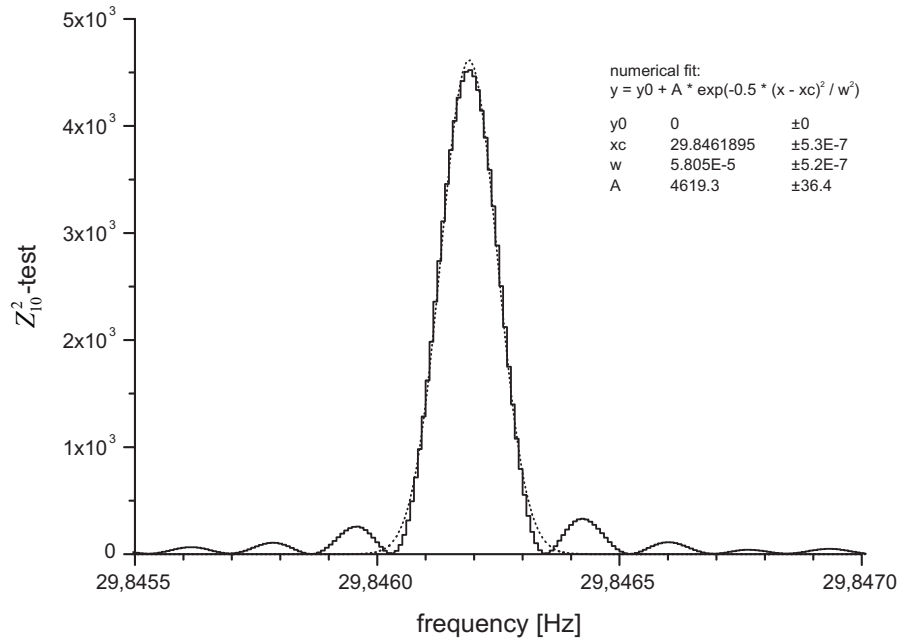, width=\textwidth}
\end{figure}

Figure~\ref{CrabPScan} shows the result of a high resolution narrow band frequency 
analysis of a ten minute OPTIMA dataset of the Crab Pulsar taken in December 1999.
Due to the relatively high brightness of the source with m$_V =~$16.6$^{m}$ the 
intensity of the detected signal is immense and the precision of the determination 
of the rotational frequency of the underlying neutron star is striking.
In figure~\ref{CrabPScan} the plotted increment in frequency between two consecutive 
calculated values of the power spectrum is 5$\cdot$10$^{-5}$ Hz (solid line) and the 
numerical parameters of the overlaid gaussian fit (dotted line) show that the 
frequency of pulsation at the observed epoch can be determined with a very small 
error of approximately 5$\cdot$10$^{-7}$ Hz \cite{Straubmeier Thesis}.

\section{Summary and Future Perspective}

With its first light observation of the Crab Pulsar in Dezember 1999 the new optical 
high-speed photometer OPTIMA of the $\gamma$-ray group of the Max-Planck-Institut 
f\"ur extraterrestrische Physik proved its scientific potential at recording the 
periodic intensity variations of the Crab Pulsar with very high time resolution.

In fall 2000 the number of operational APD detectors modules reached the proposed  
number of eight channels and OPTIMA is fully configured.
Now the possibility for a precise correction of the measured flux at the position of 
the target for the contributions of the atmospheric background and other nearby 
astronomical objects is available.
Analysing the lightcurves of periodic sources like pulsars this correction is needed 
to calculate the intensity of a possible constant emission.
Besides of that, the detector system can now determine the time resolved flux of any 
source with full correction of the sky background.
This enables us to use OPTIMA even on aperiodic targets or objects with long time 
constants.
As the slow (order of minutes) but always present changes of the atmospheric 
brightness can now be well separated from the signal of the source without the 
necessity of folding the recorded arrival-times with a short time constant, OPTIMA can 
now be used on all astronomical objects displaying fast intensity fluctuations, flares, 
eclipses or quasi-periodic oscillations. 

To demonstrate the feasibility of a precise background correction and the ability to
study a long periodic source with OPTIMA, we recorded several orbits of the eclipsing 
cataclysmic binary system HU Aquarii in summer 2000 from the 1.3m telescope on Mt. 
Skinakas, Crete.
At this time already five APD channels of the high-speed photometer were available 
and it was possible to successfully correct the flux of the binary system for the 
intensity of the atmospheric background with great accuracy.

A discussion of the OPTIMA results obtained on HU Aquarii as well as a very detailed 
analysis of our observation of the Crab Pulsar are presented in the PhD thesis of C. 
Straubmeier \cite{Straubmeier Thesis} and will be published in scientific journals in 
the near future.

~\\
Acknowledgements:
We are grateful for the provision of the Crab Pulsar ephemeris by the group of 
Andrew Lyne at Jodrell Bank. 
We would also like to acknowledge the great support we 
have received as visiting astronomers at the Calar Alto, La Silla, Mt. Stromlo and 
Skinakas observatories.
The German-Spanish Astronomical Centre, Calar Alto, is operated by the Max-Planck-Institute 
for Astronomy, Heidelberg, jointly with the Spanish National Commission for Astronomy.

\end{article}
\end{document}